\documentclass[runningheads,a4paper]{llncs}

\usepackage{amssymb}
\usepackage{graphicx}

\usepackage{url}
\usepackage[pdfdisplaydoctitle=true,
colorlinks=true,
citecolor=blue,
linkcolor=black,
urlcolor=black]{hyperref}
\newcommand{\keywords}[1]{\par\addvspace\baselineskip
\noindent\keywordname\enspace\ignorespaces#1}

\begin{document}

\mainmatter  % start of an individual contribution

\title{SADDLE: A Modular Design Automation Framework for Cluster Supercomputers and Data Centres}

\titlerunning{SADDLE: A Modular Design Automation Framework}

\author{Konstantin S. Solnushkin}
\authorrunning{Konstantin S. Solnushkin}

\institute{Saint Petersburg State Polytechnic University, Saint Petersburg, Russia\\
\email{konstantin@solnushkin.org}\\
\url{http://ClusterDesign.org/saddle}}


\toctitle{SADDLE: A Modular Design Automation Framework for Cluster Supercomputers and Data Centres}
\tocauthor{Konstantin S. Solnushkin}
\maketitle

\begin{abstract}
In this paper we present SADDLE, a modular framework for automated design of cluster supercomputers and data centres. In contrast with commonly used approaches that operate on logic gate level (Verilog, VHDL) or board level (such as EDA tools), SADDLE works at a much higher level of abstraction: its building blocks are ready-made servers, network switches, power supply systems and so on. Modular approach provides the potential to include low-level tools as elements of SADDLE's design workflow, moving towards the goal of electronic system level (ESL) design automation. Designs produced by SADDLE include project documentation items such as bills of materials and wiring diagrams, providing a formal specification of a computer system and streamlining assembly operations.

\end{abstract}

\section{Introduction and Motivation}

As of today, design automation in electronics mainly concerns using languages such as Verilog and VHDL to create devices from logic gates, or using electronic design automation (EDA) tools to create individual boards. Both approaches are well-developed, mostly because automation is indispensable in these fields. Given the current number of transistors on a chip or components on a board and the complex nature of interrelations between characteristics of systems under design, the use of computer-aided design (CAD) tools is a requirement, not a whim. Additionally, manufacture of modern electronics requires the use of industrial robots, which implies the necessity of having precise documentation that is best produced automatically with CAD tools.

On the other hand, in the field of cluster supercomputer design and data centre design in general, automation has not yet found a widespread use. An often quoted reason is that most tasks can be easily handled by a human designer, hence corresponding CAD tools do not exist because they are not required. With the size of supercomputers and data centres continuously growing, manually generated designs may no longer be cost-optimal. Common tasks that designers need to solve are: choosing types and the number of components in a server, calculating the number of servers and switches based on performance requirements, placing equipment in racks, etc. The vast size of design space justifies the use of automation.

Currently, both cluster supercomputers and warehouse-scale data centres tend to be created with identical building blocks. Deviations from established practice, if any, are infrequent and limited. We argue that this regrettable situation stems from the absence of appropriate CAD tools: when consequences of design choices are hard to predict, human designers tend to constrain the variety of their designs due to bias, personal preferences or lack of time, using familiar components in standard configurations. However, this may lead to under-exploration of the design space and thus to non-optimal designs.

Design automation tools that operate at logic gate level are still called, through inertia, ``high-level synthesis'' tools, but, compared to state-of-the-art needs, they turn out to be very low-level. The research community has called for an overarching electronic system level (ESL) design automation approach \cite{duranton2010hipeac}. SADDLE is the response to this challenge. It operates on a very high level (its building blocks are servers and switches), serving as a system-level complement to existing low-level EDA tools.

SADDLE is a modular CAD tool that allows a designer to quickly evaluate multiple design choices in terms of their technical and economic characteristics, conduct ``what-if'' scenario analyses, and automatically obtain comprehensible design documentation to simplify assembly process. The acronym stands for ``Supercomputer and data centre design language''.

\section{Related Work}

The problem of selecting an optimal configuration of a computer system has been addressed previously. One of the earliest works was \emph{R1}, an expert system created by John P. McDermott in late 1970s \cite{mcdermott1980r1}. Its main task was to configure VAX-11/780 minicomputers made by Digital Equipment Corporation. The design space was large due to an assortment of available peripheral devices; there were also various mechanical and power constraints that were taken into account. \emph{R1} was a production rule expert system which operated on a set of 480 rules representing domain knowledge. It could produce detailed assembly documentation including floor plans and cable wiring tables, and therefore set the standard for future automated configurers of computers.

In 1998, Pao-Ann Hsiung et al. \cite{hsiung1998icos} proposed \emph{ICOS}, an Intelligent Concurrent Object-Oriented Synthesis methodology which focused on design of multiprocessor systems. According to the object-oriented approach, system components are modelled as classes with hierarchical relationships between them. Previously synthesised subsystems can be reused as building blocks of new designs; machine learning and fuzzy logic are used to determine feasibility of the reuse.

In 2005, William R. Dieter and Henry G. Dietz published a technical report \cite{dieter2005automatic} detailing their methodology called \emph{Cluster Design Rules (CDR)}, as well as patterns that emerged through the continuous use of the \emph{CDR} tool. This methodology is perhaps the first attempt aimed specifically at designing cluster supercomputers. Designs were evaluated using a weighted linear combination of metrics. Although \emph{CDR} did not turn into a comprehensive product, is was a successful project that pointed to new directions for research in its field.

Among the most important observations in the practical use of the \emph{CDR} tool was that the global optimality of a supercomputer design cannot be inferred from local optimality of any of its components: for example, using a CPU with the lowest price to performance ratio does not lead to a supercomputer design with the lowest price at a given performance. This justifies thorough automated inspection of the design space.

Nagarajan Venkateswaran et al. \cite{venkateswaran2009towards} addressed the problem of automated design again in 2009. Their methodology, \emph{``Modeling and integrated design automation of supercomputers (MIDAS)''}, is aimed at Supercomputers-on-a-Chip (SCOC), but can be generalised to wider areas as well. With \emph{MIDAS}, supercom-puters-on-a-chip are built using dedicated IP cores implementing specific algorithms. The number of cores of each type to be placed on a chip is determined using performance modelling and simulation. Simulated annealing is then used to select optimal configurations. \emph{MIDAS} does not address problems of building large multi-node cluster supercomputers, but serves as an important link for implementing Electronic System Level (ESL) design approach: from chips to servers to supercomputers.

Design of large-scale computers was addressed once again by Barroso et al. in 2013 \cite{barroso2013datacenter}. However, their work mostly considers platforms for generic data centre workloads rather than high-performance computing.

SADDLE is different from the previous work in that it provides its users with a practical and useful tool for solving design problems. Its extensibility is guaranteed by modularity, allowing to interface with external design tools. Special emphasis is given to accurate estimation of economic characteristics, which is a requirement for producing cost-effective designs.

Additionally, we note that the present paper expands upon ideas previously proposed in research posters \cite{solnushkin2011combinatorial} and \cite{solnushkin2012computer}.

\section{Design Flow}
\label{Design-Flow}

Design of data centres is akin to that of cluster supercomputers, with the difference that data centres often have many smaller groups of servers, each suited for a particular task. In other respects, design for both types of installations is based on the same principles.

In SADDLE, design flow is modelled after actions of human designers. Conceptually, there are four stages: (1) exploration of compute node configurations for selection of promising candidate solutions; (2) choice of appropriate number of compute nodes in selected configuration to satisfy performance constraints; (3) automated elaboration of design decisions for infrastructural systems (network, power supply system, etc.); (4) equipment placement, floor planning and documentation generation.

From the point of view of SADDLE's user, the design procedure basically answers the question, ``What configuration of a computer has the lowest cost at a given performance?''. Configuration is defined by instances of components used to build the system and structure of their connections with each other. The detailed sequence of design steps is given below:

\begin{enumerate}
\item \label{set-of-configs} Form a set of configurations of a compute node;
\item \label{weed-out} Weed out unpromising configurations using constraints and heuristics;
\item For each configuration in the pool of candidate solutions, repeat:
	\begin{enumerate}
	\item Using inverse performance modelling, calculate the number of compute nodes required to satisfy performance constraints
	\item Design an interconnection network
	\item Design infrastructural systems (e.g., a power supply system)
	\item Place equipment into racks, locate racks on the floor, route cables
	\item Choose the best design automatically (using criterion functions) or manually
	\item Generate documentation
	\end{enumerate}
\end{enumerate}

Steps \ref{set-of-configs} and \ref{weed-out} can be repeated for as many models of servers as a designer wishes to consider, forming a bigger pool of candidate solutions for the next steps; e.g., blade servers and regular rack-mount servers can be inspected side by side, and results compared to choose the best alternative.

On every stage, semi-complete designs can be checked against a set of constraints specified by the designer; these can be physical constraints stipulated by available machine room space, power, etc., or budgetary constraints imposed on capital or operating costs. Violating configurations can be immediately discarded; this way they don't participate in further design steps which would lengthen the total time to solution.

SADDLE is modular; some of the modules are implemented as subroutine calls while others can be queried over the network.

\section{Representation of Configurations}
\label{section:Representation-of-Configurations}

Building blocks in SADDLE are compute nodes. Their configurations are represented with directed acyclic multipartite graphs: partitions correspond to functions, vertices denote components or, more generally, possible implementations of each function, while edges represent compatibility of components. Each path in the graph represents a valid configuration of a technical system, in this case, a compute node. While the use of graph theory for representing configurations is not new, a major advancement comes from assigning to vertices sets of expressions which are evaluated during graph traversal.

\begin{figure}
\centering
\includegraphics[width=90mm]{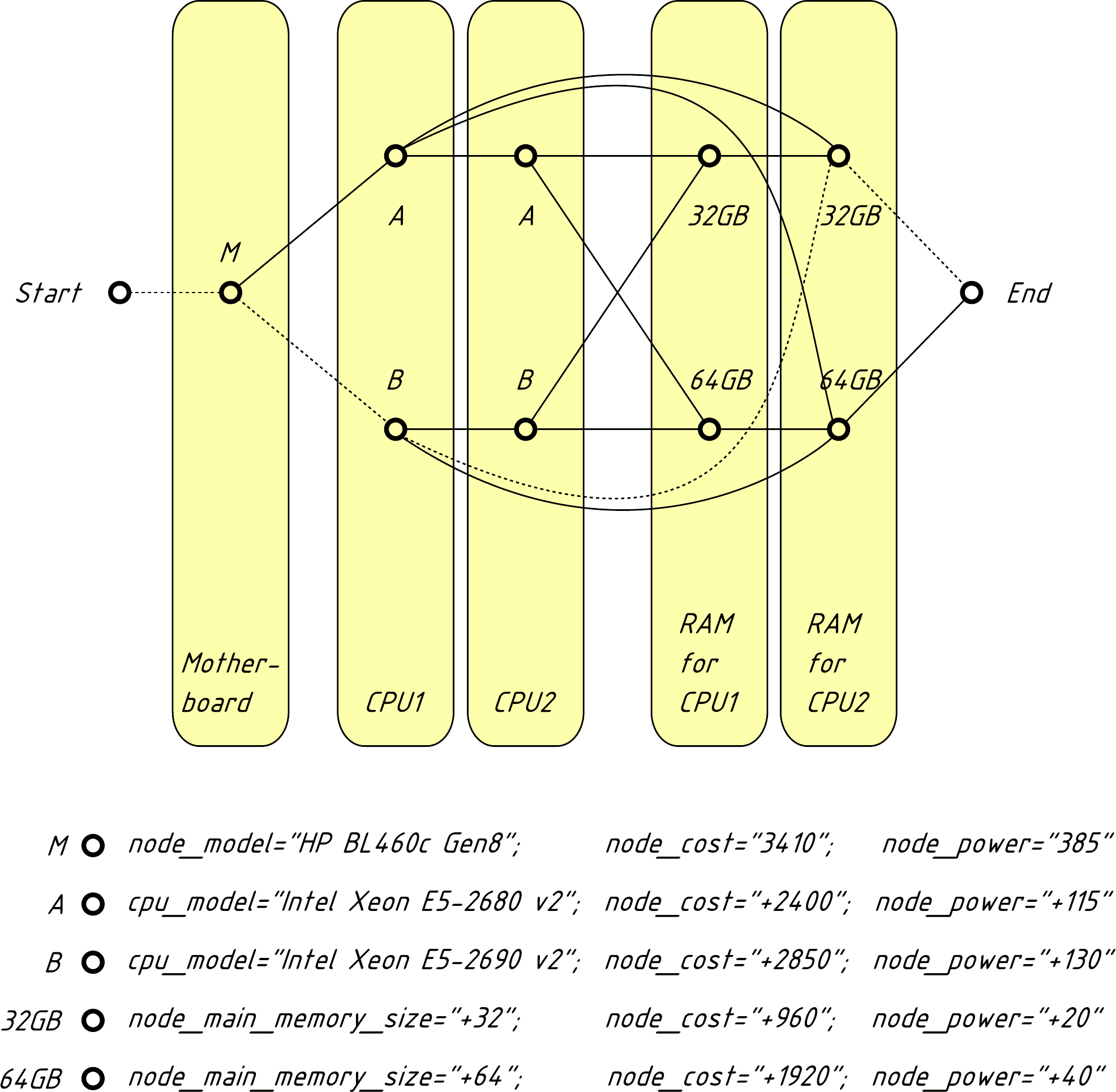}
\caption{Graph representation of eight configurations of a dual-CPU compute node.}
\label{graph-representation}
\end{figure}

For example, the graph in Figure~\ref{graph-representation} represents a dual-CPU compute node. Vertices from the database are instantiated in the graph as many times as necessary. Graph traversal yields eight paths, each corresponding to a valid configuration. Arrows on the edges are not shown to reduce clutter since traversal is performed in a natural start-to-end pattern.

One of the paths is highlighted with dotted lines. It includes vertices ``M'', ``B'' and ``32GB''; in other words, the corresponding configuration includes a motherboard, one ``Intel Xeon E5-2690~v2'' CPU and 32~GBytes of memory. Every time a vertex is traversed, expressions prescribed to it are evaluated; for example, expression \verb|node_cost="+2850"| for vertex ``B'' means to add a specified number to the existing value of this metric, or to zero if no current value exists. In this case, by the time the ``End'' vertex is reached, metric \verb|node_cost| will receive a value of ``7220''; that is the cost of this configuration of the compute node.

Virtually any technical or economic metric can be calculated in a similar manner. For example, we calculate the number of CPU cores per node, peak floating-point performance, power consumption and weight of a node, amount of RAM per core and many other metrics.

For brevity, we will omit details of expression grammar and graph transformations. We note, however, that directed acyclic graphs, when used for representing configurations, have a somewhat lower expressive power than undirected graphs with cycles, yet their visual comprehension is easier.

Graphs representing configurations of currently produced servers with real-world complexity tend to generate from 50 to 250 configurations. Most of them will not lead to good designs of cluster supercomputers and therefore can be discarded. In SADDLE, there are two mechanisms to do this: constraints and heuristics. Constraints can be specified on any metrics of a configuration. For example, the user can request to discard configurations that have CPUs with too few or too many cores, or where the amount of memory per core is too small, or configurations that don't have a specific type of network adapter.

Constraints allow to quickly reduce the size of design space, dramatically decreasing overall time to solution. However, the user should be careful not to prune too aggressively, because intuition cannot serve as a substitution for exhaustive search. For example, the user may be tempted to discard CPUs with the lower number of cores, only to find later that they actually had a larger amount of cache memory per core and would have delivered better performance for a customer's application.

As also mentioned in section~\ref{Design-Flow}, constraints can be imposed at any stage of the design process; this way, even if suboptimal configurations were not discarded immediately after generation, they can still be weeded out at later stages.

The second mechanism to deal with combinatorial explosion is heuristics. Already calculated metrics can be arbitrarily combined to form a simple objective function, and configurations can be discarded based on the value of this function. For example, we found that using the ratio of compute node's cost to its peak floating-point performance as a predictor of quality allows to safely discard 80\% of configurations.

\section{Performance Modelling}

The next design step involves calculation of the number of compute nodes required to reach a specified performance goal. In general, performance cannot be inferred from peak floating-point performance that is calculated when generating configurations; this is especially true for workloads that are not floating-point oriented. Instead, SADDLE calls design modules that implement performance models.

A performance model generally accepts on its input a number of metrics of a single compute node as well as the number of nodes (or other ``computing blocks'': cores, CPUs, accelerator boards, racks, etc.), and outputs projected performance of a cluster supercomputer. We call such models ``direct''. We also introduce the notion of \emph{inverse} performance models; these accept metrics of a compute node and desired performance, and output the number of compute nodes required to achieve the specified performance level.

Inverse performance modelling is a specific type of an inverse problem. We solve the problem in two stages: in a forward pass, we repetitively call a corresponding direct performance model with monotonically increasing the number of nodes (say, in powers of 2). When the specified performance goal has been overreached, we use bisection method to determine a more exact number of ``computing blocks''.

For demonstration purposes, we implemented a simple analytical performance model of computational fluid dynamics (CFD) software ``ANSYS/Fluent''; it's fast although not very precise. It predicts performance based on CPU clock frequency, the number of CPU cores in a cluster and the type of interconnection network (InfiniBand or 10Gbit Ethernet). Inverse performance model is implemented in the same module and calls direct performance model according to the above algorithm. We found that only a few calls to the direct model are needed to identify with sufficient precision the number of CPU cores that a cluster computer is required to have.

Generally, performance models are not limited to analytic, they can be of any nature, including simulators or FPGA prototypes, although these require more time and resources for direct and especially inverse performance modelling. This generality can be used to plug-in the results of low-level design workflow into SADDLE's high-level workflow.

Indeed, suppose there is a new CPU under development, and there exists a simulation framework that can predict performance of a cluster computer based on this type of CPU on certain workloads. Several CPU parameters can be tweaked (number of cores, sizes of caches), and each valid combination represents a separate model of this future CPU. Existing EDA tools can be used to calculate performance and power consumption of each model, with tools such as CACTI \cite{muralimanohar2007optimizing} (for memory hierarchies) and Orion \cite{kahng2009orion} (for networks-on-chip) comprising the end-to-end simulation framework. With more effort we can also estimate per-item cost of producing each CPU model when using mass production. This is enough to construct configuration graph from section~\ref{section:Representation-of-Configurations}.

Then, a simple design module can be created that queries the aforementioned simulation framework, estimating performance and cost for each cluster configuration when varying CPU model used and the number of CPUs in a cluster. This will allow engineers to use SADDLE to perform directed search of best CPU configurations, and to further build cost-effective cluster configurations. It is important to note that it is not necessarily the cheapest or the fastest CPU that brings optimality to the cluster configuration as a whole, hence the possibility to perform automated directed search is very beneficial.

Using end-to-end simulation will allow to perform comprehensive what-if analysis, answering questions such as ``How will adding more cache memory or more floating-point units to a CPU impact performance, power and cost of the whole supercomputer and its infrastructure?''

Compare this to the approach used by Sandia's SST \cite{hendry2012sst}, which creates a loop of low-level, fine-grained simulators that feed results into higher-level modules. Our approach generalises that of SST by extending it to the data centre level and by more thoroughly dealing with costs, both capital and operating.

Put another way, existing EDA tools search for a compromise between performance, power and area of an individual chip, while SADDLE searches for the same compromise for a data centre, also adding cost to the mix of metrics.

\section{Design of Subsystems}

SADDLE can design interconnection networks and power supply systems by querying design modules. Modules are web applications that can also be used standalone through a web browser. Design of other infrastructure such as storage systems can be enabled by creating corresponding modules.

SADDLE queries the network design module to design fat-tree and torus interconnection networks. Module's database contains monolithic and modular switches, using the same graph representation as detailed in section~\ref{section:Representation-of-Configurations}, with cost, power, size and weight figures for each switch configuration. As a result, all essential metrics of interconnection networks can be estimated and integrated into the overall cluster design.

Input parameters for the network design module are the number of compute nodes to be interconnected and the desired topology. The module can use any objective function; different configurations of switches are tried in an attempt to minimise this function, which by default is equipment cost.

Choosing an uninterruptable power supply (UPS) system is done by querying another design module that works very similarly to the above. In this case, input parameters are electric power that must be provided and an optional backup time when running on batteries.

\section{Equipment Placement and Floor Planning}

Inverse performance model returns the number of servers in the future supercomputer, and individual design modules return the amount of infrastructural equipment such as network switches and uninterruptable power supplies. With this information, the next stage can begin: placing equipment into racks and locating racks on the floor. SADDLE already implements a simple algorithm for these tasks, and more elaborate algorithms can be added as necessary.

Currently, SADDLE uses a set of rules to place equipment into racks. These rules are akin to heuristics proposed by Mudigonda et al. \cite{mudigonda2011taming}, but ours are more general as they allow to place equipment of different sizes. Placement occurs in the following order:

\begin{enumerate}
\item Core switches;
\item ``Compute blocks''; each block is an edge switch together with compute nodes connected to it
\item UPS equipment
\end{enumerate}

Every of these equipment types can be placed according to one of three strategies: (a) consolidation: equipment items are placed into racks as densely as possible; (b) separation: each item is placed in its own rack; (c) spread: each subsequent item is placed in a rack that is $N$ racks apart from the previous item; the latter allows to ``spread'' equipment in a machine room.

Core switches are by default placed using separation strategy. Compute blocks are placed using consolidation strategy: compute nodes from a block are added to the first rack that has enough free space, and then the accompanying edge switch is added to the same or a nearby rack. When the current rack cannot fit any more compute blocks, the next rack is used (or created, if necessary).

While edge switches should preferably be placed in the same rack with their compute nodes, they do not have to be physically adjacent. We place edge switches to the top of the rack in an effort to minimise the length of inter-rack cables, while compute nodes are placed in the bottom of the rack to ensure mechanical stability. Therefore placement results in ``gaps'' in the middle of the racks, rather than at the top or at the bottom.

The next step is to locate racks on the floor. Yet another design module is used to determine optimal dimensions of the machine room, in terms of the number of rows and racks per row, taking into account rack dimensions and clearances and trying to produce a roughly square shape.

Racks are laid out on the floor in a serpentine pattern; this is intended to ensure that a block of compute nodes in the end of a row is not located too far away from its edge switch (which is placed separately and can in principle be put in the next row).

SADDLE calculates cable lengths using Manhattan distance, prints the list of required cables and the table detailing their connections, and draws front views of racks, with cables routed in overhead trays, in SVG (Scalable Vector Graphics) format.

Cable length can be minimised by optimally locating switches in racks and racks on the floor. For example, Fujiwara et al. \cite{fujiwara2012cabinet} formulate rack layout problem as a facility location problem, where the total inter-rack cable length is sought to be minimised, and solve it using simulated annealing, reducing total cable length by 29..40\%. Similar algorithms can be added to SADDLE in the future.

\section{Experimental Evaluation}
\label{section:Experimental-Evaluation}

To evaluate SADDLE, we used it to design two cluster supercomputers, with peak floating-point performance of 100~TFLOPS and 1~PFLOPS, respectively. First, we prepared a database of compute nodes that defines the configuration graph. We used ``BL460c Gen8'' blade servers made by Hewlett-Packard, with one or two CPUs and without accelerator boards such as GPGPUs. This graph generates 56 configurations of compute nodes. We imposed constraints to select configurations with InfiniBand adapters, and then used a heuristic, ranking configurations according to the ratio of the cost of individual compute node to its peak floating-point performance and choosing a configuration with the lowest value of this metric.

This turned out to be a configuration with two ten-core Intel Xeon CPUs. We then used this configuration as a building block of our cluster supercomputers. The two designs are compared in Table~1. Operating costs were calculated with the following assumptions: system lifetime is 3 years, electricity price is \$0.35 per kW$\cdot$hour, and price of stationing one rack in a data centre are \$3,000 per year. Note that prices are retail and therefore do not reflect possible discounts for such large-scale procurements.

``Tomato equivalent'' mentioned in the last line of the table refers to the idea of reusing waste heat from the supercomputer for agricultural purposes, such as growing tomatoes in greenhouses. Calculations by Andrews and Pearce \cite{andrews2011environmental} indicate that tomato crops could reach roughly 400~kg per 1~MW of reused heat per day.

SADDLE script to produce the designs is given in Figure~2. Seventeen lines of code are enough to make quick conclusions and facilitate more detailed search. The script also highlights possibilities of performing ``what-if'' analyses: for example, settings such as electricity price, system lifetime, rack height or stationing price, etc. can all be quickly changed and the script re-run to estimate design-wide changes.

\begin{table}
\centering
\label{two-designs-table}
\caption{Comparison of two cluster computer designs, with peak floating-point performance of 100~TFLOPS and 1~PFLOPS}
\begin{tabular}{ l c c }
\hline\noalign{\smallskip}
& ~~~100~TFLOPS~~~ & ~~~1~PFLOPS~~~ \\
\noalign{\smallskip}
\hline
\noalign{\smallskip}
Compute node model & \multicolumn{2}{c}{ HP BL460c Gen8 } \\
CPU model & \multicolumn{2}{c}{ Intel Xeon E5-2680 v2 } \\
CPU clock frequency, GHz & \multicolumn{2}{c}{ 2.8 } \\
Node peak performance, GFLOPS & \multicolumn{2}{c}{ 448 } \\
Node power, W & \multicolumn{2}{c}{ 651 } \\
\noalign{\smallskip}
\hline
\noalign{\smallskip}
Number of compute nodes & 224 & 2,233 \\
Number of racks (incl. UPS) & 8 & 72 \\
Floor space size, m$^2$ & 24 & 144 \\
Cable length, m & 1,066 & 26,392 \\
Power, kW & 159 & 1,600 \\
Weight, tonnes & 3.3 & 31.6 \\
Costs, millions US dollars: & \multicolumn{2}{c}{ ~ } \\
~~~Capital expenditures & 3.1 & 31.9 \\
~~~Operating expenditures & 1.6 & 15.7 \\
~~~Total cost of ownership & 4.6 & 47.5 \\
Tomato equivalent, kg per day & 63 & 630 \\
\hline
\end{tabular}
\end{table}

\begin{figure}
\label{saddle-example-script}
\begin{verbatim}
# Open the database of configurations from specified files
open_db(['db/hp.xml', 'db/intel-xeon-2600.xml',
   'db/hp-blade-memory.xml', 'db/hp-network.xml',
   'db/hp-bl460c_gen8.xml'])
# Only allow configurations with InfiniBand connectivity
constraint("'InfiniBand' in network_tech")
# Use a heuristic to filter out unpromising configurations
metric("node_cost_to_peak_performance = node_cost /
   node_peak_performance")
# Select the best configuration according to the heuristic
select_best("node_cost_to_peak_performance")
# Delete inferior configurations
delete()
# Set the number of nodes for 1 PFLOPS of performance and update
# metrics to calculate the number of cores automatically
metric("nodes = ceil(1000000 / node_peak_performance)")
update_metrics()
# Use peak performance instead of calling a model
performance_module['url']='peak'
# Calculate performance for the said number of nodes
performance()
# Specify explicitly to use blade switches from Hewlett-Packard
metric("network_vendor='hp-blade'")
# Design a network
network()
# Design a UPS system
ups()
# Save designed equipment in a group
add_group('compute')
# Place equipment as densely as possible
place(place_params={'strategy': 'consolidate'})
# Route cables and calculate their length
cables()
# Print technical and economic metrics of the design
print_design()
# Draw and save front view of rows
draw_rows([], 'racks.svg')
\end{verbatim}
\caption{SADDLE script to produce sample designs from Table~1}
\end{figure}

\subsection{Avalanche Changes}

SADDLE also makes it easy to explore the avalanche effect of seemingly small changes on the whole design. For example, if blade servers are replaced with ordinary rack-mount servers (even with the same internal components), this immediately changes characteristics such as compute node size and cost. Different network switches then need to be used, and power consumption of the machine may also change, leading to a different configuration of the UPS system.

To explore this scenario, we designed a 1~PFLOPS machine with rack-mount servers from a different manufacturer, keeping node parameters (CPU and RAM) the same. Such compute nodes are 32\% cheaper; however, the complete machine built with them is only 18\% less expensive in terms of the total cost of ownership, because network and UPS cost do not change significantly. At the same time, this machine, due to lesser density of computing equipment, occupies 100 racks instead of 72, eventually leading to a floor space that is 30\% bigger (187~m$^2$ versus 144~m$^2$). In other words, when space is not a hard constraint, ordinary servers can be used to reduce the total cost of ownership.

\section{Extending SADDLE}

SADDLE is implemented on top of the Python ver.~3 programming language; its statements are Python functions. It is \emph{intended} to be modified and extended by its users, and its source code is heavily commented to facilitate this: about 40\% of lines are comments. Instead of being a massive piece of software with many rarely used knobs, it allows its user to make quick ad-hoc fixes for unconventional usage scenarios.

For example, operating costs are usually calculated on the assumption of continuously running hardware. If hardware only runs during part of the day, the corresponding line that calculates electricity costs can be found in the source code and edited for this particular run.

Future work is planned to further extend SADDLE. Advancements can be made to improve core functionality, for example, to enable automated design of storage and cooling systems.

Performance models, which are of great importance to SADDLE, can be created in cooperation with the authors of supercomputer software, to guarantee that SADDLE produces reliable performance estimates for various architectures, thereby allowing their fair comparison.

Another important task is to ensure that databases contain current models and prices for all types of hardware: compute nodes, network switches and UPS systems. Open format makes it easy for vendors to compile and publish databases for their hardware as often as required.

\bibliographystyle{splncs}
\bibliography{../text/bibliography/literature}

\end{document}